\documentclass[12pt]{article}
\usepackage{graphicx}
\usepackage{float}
\usepackage{hyperref}
\def\beq{\begin{equation}}
\def\eeq#1{\label{#1}\end{equation}}
\def\eeqn{\end{equation}}

\def\beqa{\begin{eqnarray}}
\def\eeqa#1{\label{#1}\end{eqnarray}}
\def\eeqan{\end{eqnarray}}

\def\overbar#1{\overline{#1}}

\let\bar=\overbar

\def\Dslash{\not{\hbox{\kern-4pt $D$}}}
\def\dslash{\not{\hbox{\kern-2pt $\del$}}}

\def\msb{{\bar{\ssstyle M \kern -1pt S}}}

\newcommand\pubnumber{SNSN-323-63}
\newcommand\pubdate{\today}

\def\strasbourg{Institut Pluridisciplinaire Hubert Curien\\
67037 Strasbourg, FRANCE}

\def\Title#1{\begin{center} {\Large #1 } \end{center}}
\def\Author#1{\begin{center}{ \sc #1} \end{center}}
\def\Address#1{\begin{center}{ \it #1} \end{center}}

\newcommand\pubblock{\rightline{\begin{tabular}{l} \pubnumber\\
         \pubdate  \end{tabular}}}
\newenvironment{Abstract}{\begin{quotation}  }{\end{quotation}}
\newenvironment{Presented}{\begin{quotation} \begin{center} 
             PRESENTED AT\end{center}\bigskip 
      \begin{center}\begin{large}}{\end{large}\end{center} \end{quotation}}

\def\beq{\begin{equation}}
\def\eeq#1{\label{#1}\end{equation}}
\def\eeqn{\end{equation}}

\def\beqa{\begin{eqnarray}}
\def\eeqa#1{\label{#1}\end{eqnarray}}
\def\eeqan{\end{eqnarray}}

\def\overbar#1{\overline{#1}}

\let\bar=\overbar

\def\Dslash{\not{\hbox{\kern-4pt $D$}}}
\def\dslash{\not{\hbox{\kern-2pt $\del$}}}

\def\msb{{\bar{\ssstyle M \kern -1pt S}}}

\begin{document}

\begin{titlepage}
\pubblock

\vfill
\Title{Measurement of the tZq cross section at 13 TeV with the CMS detector}
\vfill
\Author{Nicolas Tonon on behalf of the CMS Collaboration}
\Address{\strasbourg}
\vfill
\begin{Abstract}
Evidence is presented for standard model production of a Z boson in association with a single top quark. The analysis uses a data sample of proton-proton collisions at $\sqrt{s} = 13$ TeV recorded in 2016 by the CMS detector, corresponding to an integrated luminosity of 35.9 fb$^{-1}$. Final states containing three leptons (electrons or muons) and at least two jets are investigated, and the measured cross section is $\sigma($pp $\rightarrow$ tZq $\rightarrow $Wb$\ell^{+}\ell^{-}$q) = 123$^{+44}_{-39}$ fb, where $\ell$ stands for electrons, muons or taus. The corresponding observed (expected) significance is 3.7 (3.1) standard deviations.
\end{Abstract}
\vfill
\begin{Presented}
TOP 2017 Conference\\
Braga, Portugal,  September 17--22, 2017
\end{Presented}
\vfill
\end{titlepage}
\def\thefootnote{\fnsymbol{footnote}}
\setcounter{footnote}{0}

\section{Introduction}

%\paragraph{}
At the CERN LHC, top quarks are predominantly produced in pairs ($t \overbar t$) via the strong interaction. Single top quarks can also be produced via the electroweak interaction. The unprecedented centre-of-mass energy reached at the LHC and its large luminosity make it possible to study very rare processes for the first time. This analysis aims at measuring the cross section of one such process: the associated production of a single top quark and a Z boson, denoted tZq.
For proton-proton collisions at a centre-of-mass energy of 13 TeV, the cross section of the tZq process considering only the leptonic decays of the Z boson (to electrons, muons or taus), including lepton pairs from off-shell Z and $\gamma$ bosons with invariant mass $m_{\ell^{+}\ell^{-}} > 30$ GeV, is calculated\footnote{Using MC@NLO \cite{NLO} at next-to-leading order in the five-flavour scheme.} to be $\sigma^{SM}($tZq $\rightarrow$ t$\ell^{+}\ell^{-}$q) = 94.2$^{+1.9}_{-1.8}$ (scale) $\pm$ $2.5$ (PDF) fb. A previous search at 8 TeV with the CMS experiment \cite{CMS} found an observed significance of 2.4 standard deviations \cite{CMS8tev}. The ATLAS Collaboration recently reported evidence for tZq production at 13 TeV \cite{ATLAS}.  \\
%\paragraph{}
The observation and subsequent measurement of the cross section of this production mechanism allow both to probe the Standard Model (SM) in a unique way and to help constrain beyond-SM models.
For example this process is sensitive to the top quark coupling to the Z boson, but also to the triple gauge-boson coupling WWZ.
Moreover, tZq is sensitive to flavour-changing neutral current (FCNC) processes which share a similar final state. These processes are strongly suppressed within the SM due to the GIM mechanism \cite{GIM}, but are predicted to be largely enhanced in several beyond-SM models. Therefore, deviations from the expected SM tZq production cross section could be indicative of the presence of new physics.

\section{Event selection and background estimation}

%\paragraph{}
The signature of the tZq process consists of a single top quark, a Z boson and an additional (``recoiling'') jet emitted in the forward region of the detector (absolute pseudorapidity $|\eta| > 2.4$). This analysis targets events where both the Z boson and the W boson coming from the top quark (t $\rightarrow$ Wb) decay leptonically (either to electrons or muons), leading to a final state containing three leptons, a light and bottom quark, and missing transverse energy arising from the undetected neutrino. This leads to four possible final states depending on the leptons flavours : $\mu\mu\mu$, $\mu\mu$e, ee$\mu$ and eee. \\
%\paragraph{}
All selected events are required to contain exactly three isolated leptons with transverse momentum $p_{T}$ greater than 25 GeV. Electrons (muons) must satisfy $| \eta |$ smaller than 2.5 (2.4). Among the three selected leptons, there must be an opposite-sign same-flavour pair having an invariant mass compatible with that of a Z boson ( $76 < m_{\ell^{+}\ell^{-}} < 106$ GeV). 
%The additional lepton is associated to the top quark decay. 
Jets are required to satisfy $p_{T} > 30$ GeV and must be within the extended pseudorapidity range $| \eta | < 4.5$, to account for the forward jet. 
Jets that originate from the hadronisation of a b quark are identified (tagged) using the combined secondary vertex (CSVv2) algorithm \cite{btag1} \cite{btag2}, which combines various track-based variables with secondary-vertex variables to construct a discriminating observable in the region $| \eta |$ $<2.4$. The chosen working point of the algorithm corresponds to an efficiency $\epsilon_{btag} = 83 \%$ with a mistag rate $\epsilon_{mis} = 10 \%$.
To reduce the contributions from backgrounds containing four leptons in the final state (e.g. ZZ, $t\overbar tZ$ and $t \overbar tH$), events containing an additional fourth lepton with $p_{T}>10$ GeV are not considered.\\
%\paragraph{}
The final state targeted in this analysis contains sizeable contributions from backgrounds containing three prompt leptons (from W or Z decay), mainly WZ+jets and $t\overbar tZ$. There is also a large contamination from backgrounds containing two prompt leptons and one \textit{not-prompt} lepton, which is either coming from heavy-flavour hadron decay, from photon conversion, or was misreconstructed as a lepton. This contamination is almost entirely due to the $t \overbar t$ and Drell-Yan (DY) processes, which are referred to as the ``not-prompt background''. Unlike all other backgrounds, this sample is evaluated directly from the data, as these processes are known to be poorly modeled by the Monte Carlo simulation. To obtain it, the isolation requirement is reverted and the identification criterion is loosened for one lepton among any of the three. This procedure ensures that the correct fractions for both processes ($t\overbar t$ and DY) are correctly represented in the final sample.\\
%\paragraph{}
Three orthogonal regions are then defined based on the jet and b-jet multiplicities. The ``1bjet'' signal region is defined as containing two or three jets, among which exactly one b-tagged jet. The ``2bjets'' control region, which also contains some signal and is useful to control the $t\overbar tZ$ background, contains events with at least 2 jets among which at least two b-jets. Finally the ``0bjet'' control region, enriched in WZ+jets and not-prompt backgrounds, contains events with at least one jet among which no b-jet. A preliminar fit to the data is performed in the 0bjet region, with only the not-prompt contribution let floating, in order to normalise this data-driven sample.

\section{Analysis strategy and results}

\begin{figure}[H]
  \begin{center}
       %\resizebox{10cm}{!}{\includegraphics{images/postfit_input_tZq_ttZ_all}}
       \includegraphics[width=12cm, height = 6.5cm]{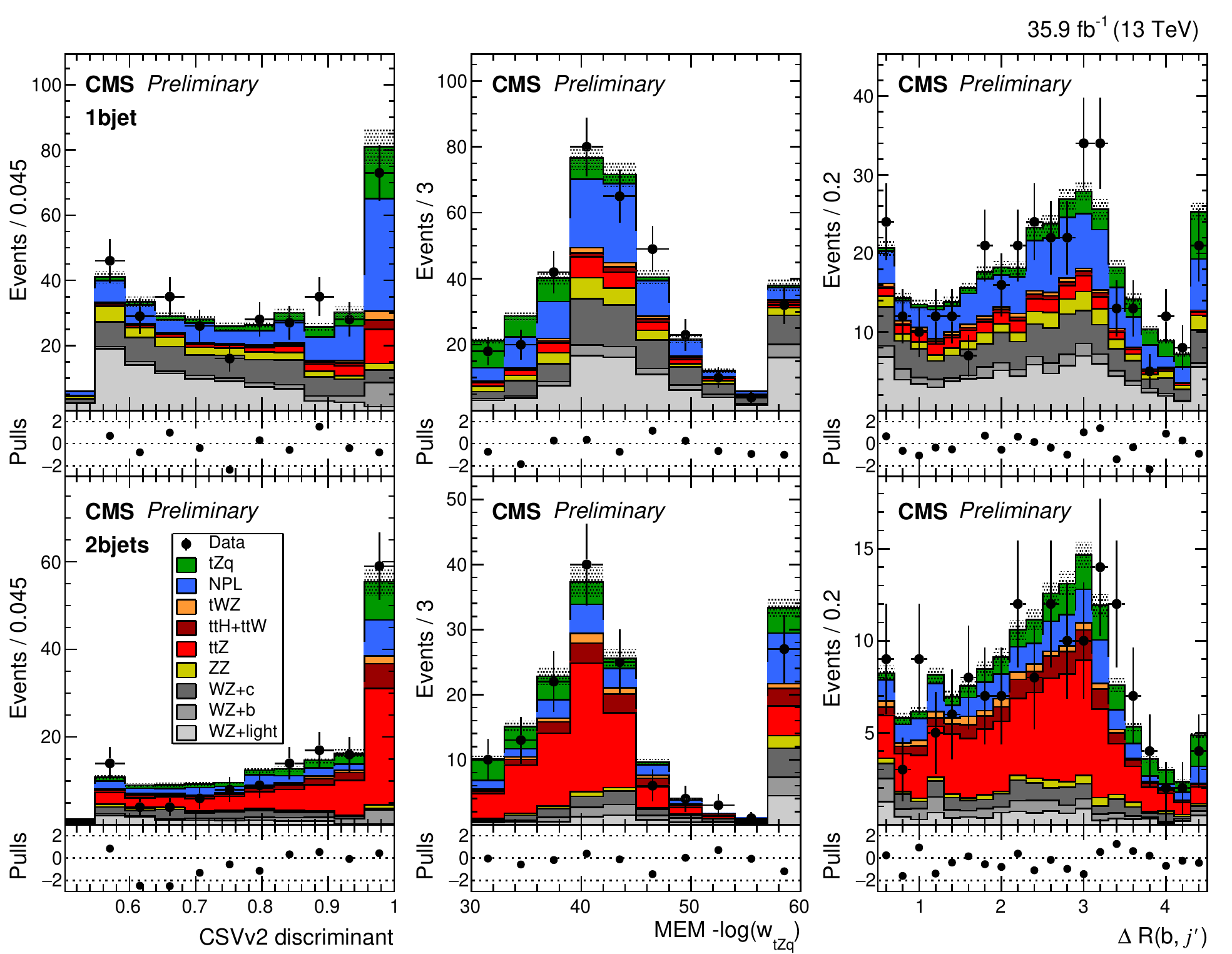}
  \end{center}
    \caption{Data-to-prediction comparisons in the 1bjet region (upper row) and in the 2bjets region (bottom row) for the largest CSVv2 discriminant value among all selected jets (left), the logarithm of the MEM score associated to the most probable tZq kinematic configuration (centre), and the $\Delta R$ separation between the b quark and the recoiling jet (right). The distributions include events from all final states. Underflows and overflows are shown in the first and last bins, respectively. The hatched bands include the total uncertainty on the background and signal contributions. The pulls in the distributions are shown in the bottom panels. Figure taken from \cite{PAS}.}
    \label{fig:tZq}
\end{figure}

%\paragraph{}
In the 1bjet and 2bjets regions, a multivariate approach is implemented. Boosted Decision Trees (BDT) are trained to discriminate the signal against the $t \overbar tZ$, WZ+jets and ZZ backgrounds. While most of the input variables are kinematic and angular observables, a few variables (four in the 1bjet region, two in the 2bjets region) are obtained with the Matrix Element Method (MEM) \cite{MEM}. 
The principle of this method is to compute a weight for each event (both for data and simulation) depending on a given hypothesis. In this analysis, the weights are computed under the signal, $t\overbar tZ$ and WZ+jets hypotheses. Therefore this method provides variables which quantify the compatibility of events with each of these processes. These variables are then included as any other to train the BDTs. 
The addition of these high-level variables are found to significantly improve the discrimination powers of the BDTs.\\
%\paragraph{}
The signal is extracted by performing a likelihood fit simultaneously in the three regions and in the four channels. The not-prompt background normalisation is let free in the fit. In the 0bjet control region, the distribution of the transverse mass of the W boson is used. It is defined as $m_{T}^{W} = \sqrt{2 \, p_{T} \, p_{T}^{miss} \, (1- \cos(\Delta \phi))}$, where $p_{T}$ is the transverse momentum of the lepton produced in the W boson decay and $\Delta \phi$ is the difference in azimuth between the direction
of the lepton and the direction of $p_{T}^{miss}$.
In the two other regions, the distributions of the BDT discriminants are used. Data-to-prediction comparisons in the 1bjet and 2bjets regions are shown for three of the most discriminating input variables in Fig. \ref{fig:tZq}.\\
%\paragraph{}
The cross section is measured to be $\sigma($tZq $\rightarrow$ t$\ell^{+}\ell^{-}$q) = 123$^{+33}_{-31}$ (stat.) $^{+29}_{-23}$ (syst.) fb, which means that an excess of events is observed which is compatible with the Standard Model prediction within one standard deviation. The corresponding observed and expected significances are 3.7 and 3.1 standard deviations respectively.

%%%%%%%%%%%%%%%%%%%%%%%%%%%%%%%%%%%%%%%%%%%%%%%%%%%%%%%%%%%%%%%%%%%%%%%%%
%%
%%   use this format to include an .eps figure into your paper
%%

%%%%%%%%%%%%%%%%%%%%%%%%%%%%%%%%%%%%%%%%%%%%%%%%%%%%%%%%%%%%%%%%%%%%%%%%%%%

%\bibliography{TOP2017_proc_tonon}

\begin{thebibliography}{99}

%%
%%  bibliographic items can be constructed using the LaTeX format in SPIRES:
%%    see    http://www.slac.stanford.edu/spires/hep/latex.html
%%  SPIRES will also supply the CITATION line information; please include it.
%%

\bibitem{CMS}
%\href{https://cds.cern.ch/record/2138504}{CMS Collaboration, ``The CMS experiment at the CERN LHC'', JINST 3 (2008) S08004, doi:10.1088/1748-0221/3/08/S08004.}
CMS Collaboration, ``The CMS experiment at the CERN LHC'', JINST 3 (2008) S08004, doi:10.1088/1748-0221/3/08/S08004.
\

\bibitem{CMS8tev}
CMS Collaboration, ``Search for associated production of a Z boson with a single top quark and for tZ flavour-changing interactions in pp collisions at $\sqrt{s}=8$ TeV'', JHEP 07 (2017) 003, doi:10.1007/JHEP07(2017)003.
\

\bibitem{ATLAS}
ATLAS Collaboration, ``Measurement of the production cross-section of a single top quark in association with a Z boson in proton-proton collisions at 13 TeV with the ATLAS detector'', Submitted to Phys. Lett. B (2017), arXiv:1710.03659.
\

\bibitem{NLO}
J. Alwall et al., ``The automated computation of tree-level and next-to-leading order differential cross sections, and their matching to parton shower simulations'', JHEP 07 (2014) 079, doi:10.1007/JHEP07(2014)079.
\


\bibitem{GIM}
S. Glashow, J. Iliopoulos, and L. Maiani, ``Weak Interactions with Lepton Hadron
Symmetry'', Phys. Rev. D 2 (1970) 1285, doi:10.1103/PhysRevD.2.1285.
\

\bibitem{btag1}
CMS Collaboration, ``Identification of b-quark jets with the CMS experiment'', JINST 8 (2013) P04013, doi:10.1088/1748-0221/8/04/P04013/
\

\bibitem{btag2}
CMS Collaboration, ``Identification of b quark jets at the CMS Experiment in the LHC Run 2'', CMS-PAS-BTV-15-001,
\href{https://cds.cern.ch/record/2138504}{https://cds.cern.ch/record/2138504}.
\

\bibitem{MEM}
D0 Collaboration, ``A precision measurement of the mass of the top quark'', Nature 429 (2004) 638, doi:10.1038/nature02589.
\

\bibitem{PAS}
CMS Collaboration, ``Evidence for the standard model production of a Z boson and a single top quark in pp collisions at $\sqrt{s}$ = 13 TeV'', CMS-PAS TOP-16-020,
\href{https://cds.cern.ch/record/2284830}{https://cds.cern.ch/record/2284830}.
\


\end{thebibliography}
%\bibliographystyle{Abbie}
%\bibliographystyle{plainurl}

\end{document}